\documentclass[12pt]{article}
\usepackage{amscd,amssymb,amsmath,latexsym,enumerate}
\usepackage[mathscr]{euscript}
\usepackage{epsfig}
\usepackage{fancybox}
\usepackage{verbatim}
\usepackage{graphicx}

\usepackage[normalem]{ulem}
\usepackage{multirow}
\usepackage{stackengine}

\usepackage{color}

\title{Chern numbers as half-signature of the spectral localizer}

\author{Edgar Lozano Viesca$^1$, Jonas Schober$^2$, Hermann Schulz-Baldes$^2$
\\
\\
{\small $^1$ Instituto de Matem\'aticas, UNAM, Unidad Cuernavaca, Mexico}
\\
{\small $^2$ Department Mathematik, Friedrich-Alexander-Universit\"at Erlangen-N\"urnberg, Germany}
}

\date{ }

\newtheorem{theo}{Theorem}

\newtheorem{lemma}{Lemma}
\newtheorem{coro}{Corollary}

\newcommand{\CM}{{\mathbb C}}

\newcommand{\RM}{{\mathbb R}}

\newcommand{\ZM}{{\mathbb Z}}
\newcommand{\PM}{{\mathbb P}}

\newcommand{\Hh}{{\cal H}}

\newcommand{\one}{{\bf 1}}

\newcommand{\Tr}{\mbox{\rm Tr}}

\newcommand{\SF}{{\rm Sf}} 
 
\newcommand{\Ch}{{\rm Ch}} 
\newcommand{\Ind}{{\rm Ind}} 
\newcommand{\Ker}{{\rm Ker}} 
\newcommand{\Ran}{{\rm Ran}} 
 
\newcommand{\Sig}{{\rm Sig}} 
\newcommand{\diag}{{\rm diag}}

\newcommand{\od}{{\mbox{\rm\tiny od}}}
\newcommand{\eve}{{\mbox{\rm\tiny ev}}}

\newcommand{\Tune}{\kappa}

\newcommand{\Radius}{\rho}

\begin{document}

\maketitle

\begin{abstract}
Two recent papers proved that complex index pairings can be calculated as the half-signature of a finite dimensional matrix, called the spectral localizer. This paper contains a new proof of this connection for even index pairings based on a spectral flow argument. It also provides a numerical study of the spectral gap and the half-signature of the spectral localizer for a typical two-dimensional disordered topological insulator in the regime of a mobility gap at the Fermi energy. This regime is not covered by the above mathematical results (which suppose a bulk gap), but nevertheless the half-signature of the spectral localizer is a clear indicator of a topological phase.

\vspace{.1cm}



\end{abstract}


\vspace{1cm}

\section{Introduction}
\label{sec-Intro}

In non-commutative geometry, an index pairing results from pairing a $K$-theory class with a Fredholm module \cite{Con,GVF}. In complex $K$-theory and $K$-homology, such pairings can be odd or even. Here the focus is on those pairings which lead to Fredholm operators in the classical sense and hence an integer-valued index. In two recent papers \cite{LS,LS2}, Loring and one of the authors proved that both odd and even index pairings can be calculated as the half-signature of a certain matrix, called the spectral localizer. In the even case,  the normality of the Dirac operators is needed as a supplementary condition. While the initial proofs in \cite{LS,LS2} use several $K$-theoretic tools and are quite involved, the paper \cite{LS3} gave a relatively elementary proof for the case of odd pairings. It is merely based on basic properties of the spectral flow. In the present paper, this is also achieved for the even index pairings. As will be discussed below, this provides new insights on the nature of the spectral localizer itself. We furthermore expect these proofs via spectral flow also to transpose to semifinite index pairings which could then be calculated in terms of a Breuer-Fredholm signature.

\vspace{.2cm}

While the spectral localizer is certainly of interest from a purely index-theoretic point of view, its main use may be to make index pairings accessible to numerical computations in situations where other standard tools do not work as well. This was demonstrated already in an early study by Loring \cite{Lor} and quite impressively in a recent preprint addressing quasicrystals \cite{Lor2}. In this paper the more standard situation of a two-dimensional disordered integer quantum Hall state is analyzed numerically. Particular focus is on the mobility gap regime in which the Hamiltonian has Anderson localized states at the Fermi level, but the Fermi projection nevertheless has non-trivial topology. The prototypical example is the quantum Hall regime between two Landau bands where the density of states is strictly positive and the Chern number (that is, the Hall conductivity) is non-vanishing. This Chern number is equal to an index pairing by a well-known index theorem which extends into the mobility gap regime, namely both the Chern numbers and the integer-valued index remain well-defined and are equal \cite{BES,PS}. At least in the regime of a gapped Hamiltonian, the Chern number can hence be calculated as the half-signature of the spectral localizer due to the result of \cite{LS2}. It is, of course, interesting to analyze the spectral localizer in the Anderson localized mobility gap regime.

\vspace{.2cm}

To investigate this point numerically, we picked the $p+ip$ wave dirty superconductor in the tight-binding approximation as a toy model. Even though the numerical methods are not nearly as sophisticated as in \cite{Lor2} and the used computational power was quite limited,  the first numerical results presented here are encouraging. Indeed, they show that there is large range of parameters (of the disorder strength) for which the Hamiltonian has (presumably Anderson localized) spectrum at the Fermi level and the spectral localizer remains nevertheless gapped and has a non-vanishing signature. Hence the conclusion from this numerical study is that the spectral localizer continues to detect the non-trivial topology in the mobility gap regime, even though the mathematical theorems of \cite{LS,LS2} do not apply any longer.

\vspace{.2cm}

Let us comment that also $\ZM_2$-valued (strong) index pairings make sense in the mobility gap regime whenever a Real symmetry is present \cite{Sch,GS}. It was argued in \cite{LS} that also in these cases one can calculate the $\ZM_2$-invariant as the sign of the Pfaffian or determinant of the spectral localizer (depending on the symmetry considered). Such Real symmetries are not studied in the present work. Finally let us indicate a further potential application of the spectral localizer: it can be used as a local topological marker allowing to distinguish spacial regions with different topological invariants in an inhomogeneous material or model. 

\vspace{.2cm}

This paper is organized as follows. After the definition of index pairings and the spectral localizer in Sections~\ref{sec-FredMod} and \ref{sec-SpecLoc} respectively,  the main known facts about the connection between index pairings and the spectral localizer are presented in Section~\ref{sec-Gap}. In Section~\ref{sec-Chern} it is discussed how to use the spectral localizer for the calculation of Chern numbers. The numerical results on the mobility gap regime are then presented and discussed in Section~\ref{sec-p+ip}. An outline of the new spectral flow proof of the main result on even index pairings is given in Section~\ref{sec-SFargument}, and the details of the proof follow in Section~\ref{sec-Proof}.

\section{Fredholm modules and index pairings}
\label{sec-FredMod}

Let us begin by describing the notations and notions of complex index pairings ({\it i.e.} without Real symmetries), by extracting the essentials from the general framework of non-commutative index theory \cite{Con,GVF}. Let $A$ be an invertible operator on a separable Hilbert space $\Hh$. Its phase $U=A|A|^{-1}$ is a unitary operator. Even though this is not of importance in the following, one may consider it as specifying a $K_1$-class (of the $C^*$-algebra generated by $U$ or some larger $C^*$-algebra). An unbounded odd Fredholm module for $A$ is a selfadjoint invertible (Dirac) operator $D$ with compact resolvent such that the commutator $[A,D]$ extends to a bounded operator (more traditionally, one requires commutators for a dense subset in a C$^*$-algebra containing $A$ to have this property). Associated to $D$ is the so-called Hardy projection $\Pi=\chi(D>0)$ where $\chi$ denotes the indicator function.  Then it is well-known, {\it e.g.}\ \cite{Con} or p.\ 462 in \cite{GVF}, that the commutator $[\Pi,A]$ is compact and the Toeplitz operator
\begin{equation}
\label{eq-OddPair}
T^\od\;=\;\Pi \,A\,\Pi+(\one-\Pi)
\;,
\end{equation}
is a bounded Fredholm operator on $\Hh$. To any Fredholm operator $T$ is associated its index 
$$
\Ind(T)
\;=\;
\dim(\Ker(T))\,-\,\dim(\Ker(T^*))
\;.
$$
(It should be called a Noether index rather than a Fredholm index as Fritz Noether was the first to exhibit a Fredholm operator with non-vanishing index, and Fredholm erroneously believed that all Fredholm operators have vanishing index.) The operator $T^\od$ and its index are called the odd index pairing of (the $K_1$-class of) $A$ with (the odd Fredholm module specified by) $D$.

\vspace{.2cm}

Next let us describe even index pairings. Let $H=H^*$ be an invertible selfadjoint operator on a Hilbert space $\Hh$. By spectral calculus it has a negative spectral projection $P=\chi(H<0)$ (the so-called Fermi projection of $H$) which may be thought of as fixing a $K_0$-class (of a suitable $C^*$-algebra, but again this is not relevant for the following). An even Fredholm module for $H$ is an invertible, selfadjoint (Dirac) operator $D$ on $\Hh\oplus\Hh$ with compact resolvent such that $[D,H \oplus H]$ can be extended to a bounded operator together with a selfadjoint unitary $\Gamma$ with two infinite dimensional eigenspaces and for which $\Gamma D\Gamma=-D$. In the following, we will always go into the spectral representation of $\Gamma$ so that
$$
\Gamma
\;=\;
\begin{pmatrix} \one & 0 \\ 0 & -\one \end{pmatrix} 
\;,
\qquad
D\;=\;\begin{pmatrix} 0 & D_0^* \\ D_0 & 0\end{pmatrix}
\;,
$$
where $D_0$ is an invertible, unbounded operator on $\Hh$. Furthermore, it will always be assumed below that $D_0$ is normal. The operator $F=D_0 |D_0|^{-1}$ is unitary and is called the corresponding Dirac phase. Again it is well-known \cite{Con,GVF} that 
\begin{equation}
\label{eq-EvenPair}
T^\eve\;=\; P F P \,+\, ( \one -P )
\end{equation}
is a bounded Fredholm operator which together with its index is called the even index pairing of $H$ with $D$ (or more conventionally, the $K_0$-class of $P$ with the even Fredholm module specified by $D$).

\section{Spectral localizer}
\label{sec-SpecLoc}

In this section, the spectral localizer is introduced, separately for the odd and even cases. It is a selfadjoint operator on $\Hh\oplus\Hh$ depending on a tuning parameter $\kappa>0$. For an odd pairing, it is given by
$$
L^\od_\kappa 
\;=\;
 \begin{pmatrix} \kappa\,D & A \\ A^* & -\kappa\,D\end{pmatrix}
 \;=\;
\kappa\,D\otimes\Gamma\,+\, \begin{pmatrix} 0 & A \\ A^* & 0\end{pmatrix}
\;,
$$
while for an even pairing
$$
L^\eve_\kappa 
\;=\; 
\begin{pmatrix} -H & \kappa \,D_0^* \\ \kappa\, D_0 & H\end{pmatrix} 
\;=\; \kappa \,D\,-\, H \otimes \Gamma
\;.
$$
The tuning parameter can be thought of as the resolution of space which allows to alter the distance between eigenvalues of $D$ that are interpreted as spacial distance. Note that with this interpretation in mind, the fact that the commutators of $A$ and $H$ with $D$ are bounded reflects that $A$ and $H$ are local operators w.r.t. the spacial structure of $D$, that is, their matrix elements have an off-diagonal decay over the eigenbasis of $D$. In both the odd and even case, the spectral localizer is next restricted to finite volume, again with a notion of space connected to the Dirac operator. Hence, as finite volume one uses the spectral projections of $D$ on all eigenvalue of modulus less than $\rho>0$. For odd pairings, let $\pi_\rho$ the surjective partial isometry onto  $\Hh_\rho=\Ran(\chi(|D|\leq \rho))$ which by the compactness assumption on the resolvent of $D$ is a finite dimensional subspace. Then set $(\Hh\oplus\Hh)_\rho=\Hh_\rho\oplus\Hh_\rho$ and let us identify $\pi_\rho\oplus\pi_\rho$ with $\pi_\rho$ for sake of notational simplicity. Then for any operator $B$ on $\Hh$ or $\Hh\oplus\Hh$ let $B_\rho=\pi_\rho B\pi_\rho^*$ be the restriction of $B$ to $\Hh_\rho$ or $(\Hh\oplus\Hh)_\rho$, respectively. In particular, $\one_\rho=\pi_\rho \pi_\rho^*$ is the identity on $\Hh_\rho$ or $(\Hh\oplus\Hh)_\rho$. The finite volume spectral localizer is then the finite-dimensional selfadjoint matrix
$$
L^\od_{\kappa,\rho}\;=\;(L^\od_\kappa)_\rho
\;=\;
 \begin{pmatrix} \kappa\,D_\rho & A^*_\rho \\ A_\rho & -\kappa\,D_\rho\end{pmatrix}
\;.
$$
In the case of even index pairings, one proceeds in a similar manner, but now $\Ran(\chi(|D|\leq \rho))$ is a subspace $(\Hh\oplus\Hh)_\rho$ of $\Hh\oplus\Hh$. As $D^2 = \diag (D_0^*D_0, D_0 D_0^*)$ one has $(\mathcal H \oplus \mathcal H )_\rho = \mathcal H_{\rho,+} \oplus \mathcal H_{\rho,-}$ with $\mathcal H_{\rho,+} = \Ran ( \chi(|D_0| \leq \rho))$ and $\mathcal H_{\rho,-} = \Ran ( \chi(|D_0^*| \leq \rho))$. As already stressed above, it will assumed throughout that $D_0$ is normal. Then $\Hh_{\rho,+}=\Hh_{\rho,-}$ which will again simply be denoted by $\Hh_\rho$. Then the set-up is exactly as in the case of odd pairings and the spectral localizer at finite volume is given by the same formula as above:
\begin{equation}
\label{eq-SLEvDef}
L^\eve_{\kappa, \rho} 
\;=\; \left( \kappa\, D \,-\, H \otimes \Gamma \right)_\rho 
\;=\; \begin{pmatrix} -H_\rho & \kappa \,D_{0,\rho}^* \\ \kappa\, D_{0,\rho} & H_\rho \end{pmatrix}
\;.
\end{equation}
Whenever a statement below holds for both $L^\od_{\kappa, \rho}$ and $ L^\eve_{\kappa, \rho}$, the upper index is dropped. 

\vspace{.2cm}

Before going on with the presentation of results, let us put forward some intuition on the spectral localizer.  If $A$ and $H$ vanish (what is strictly speaking not allowed), then the spectrum of $L_{\kappa}$ is symmetric around $0$ by construction. Note that the distance of the spectrum to $0$ is of the order $\kappa$ which can thus be made small by increasing the spacial resolution. Now $A$ and $H$ act like a mass term and open a larger spectral gap of the spectral localizer, by moving low-lying eigenvalues of $L_{\kappa}$ away from $0$. It is, however, a fact proved later on that this gap opening by adding $A$ or $H$ happens in a non-trivial manner, namely there may be more eigenvalues moving to the right or left of $0$. This can hence create a spectral asymmetry which turns out to be dictated by the topology captured by the index pairing. Resuming, $A$ and $H$ are like mass terms, albeit topologically non-trivial ones. Now as $A$ and $H$ are local, this spectral asymmetry of $L_\kappa$ should be captured already by its low-lying spectrum, namely it can be read off from the signature of the finite volume restrictions $L_{\kappa,\rho}$. To prove the validity of these heuristics is the object of \cite{LS,LS2,LS3}, and also the present paper.

\section{Spectral gap and half-signature of spectral localizer}
\label{sec-Gap}

\begin{theo}
\label{theo-Gap}
Let $g$ be the invertibility gap of $A$ or $H$, namely $g=\|A^{-1}\|^{-1}$ or $g=\|H^{-1}\|^{-1}$ respectively. Suppose that the tuning parameter satisfies in the respective cases
\begin{equation}
\label{eq:kappa}
\kappa 
\;\leq \;
\frac{g^3}{12 \left\|H\right\| \left\|\left[D,A\right]\right\|}
\;,
\qquad
\kappa 
\;\leq \;
\frac{g^3}{12 \left\|H\right\| \left\|\left[D,H \oplus H\right]\right\|}
\;,
\end{equation}
and that the radius $\rho$ satisfies
\begin{equation}
\label{eq:rho}
\rho\;>\;\frac{2g}{\kappa}
\;.
\end{equation}
Then
\begin{equation}
\label{eq-LBound}
(L_{\kappa, \rho})^2 
\;\geq \;
\frac{g^2}{4} \,\one _\rho
\;.
\end{equation}
In particular,  \eqref {eq:kappa} and \eqref{eq:rho} imply that $L_{\kappa, \rho}$ is invertible.
\end{theo}

The proof of this statement is given in full detail in \cite{LS2,LS3} and will not be reproduced here. Let us merely sketch the main idea, focussing on the case of an odd pairing. One starts out from 
\begin{equation}
\label{eq-LowBound}
(L^\od_{\Tune,\Radius})^2
\;=\;
\begin{pmatrix} A_\Radius^* A_\Radius & 0 \\ 0 & A_\Radius A_\Radius^*\end{pmatrix}
\,+\,
\Tune^2\begin{pmatrix} D^2_\Radius & 0 \\ 0 & D^2_\Radius \end{pmatrix}
\,+\,
\Tune \begin{pmatrix} 0 & [D,A]^*_\Radius  \\ [D,A]_\Radius  & 0 \end{pmatrix}
\;.
\end{equation}
The first two summands are positive. The second one is large on large eigenvalues of $D$, but relatively small for on the low-lying spectrum of $D$ due to the (small) factor $\kappa^2$. For the latter spacial region, the positivity comes from the positivity of the first summand. Now $(A^*A)_\rho$ is bounded below by $g^2\one_\rho$, but $A_\Radius^* A_\Radius\not= (A^*A)_\rho$. On the other hand, one can use an operator $f_\rho=f_\rho(D)$ constructed from a tapering (smooth) function $f_\rho:[-\rho,\rho]\to[0,1]$ of $D$ which vanishes at $\pm\rho$ and is equal to $1$ on $[-\frac{\rho}{2},\frac{\rho}{2}]$. Then 
$$
A_\rho^*A_\rho
\;\geq\; 
\pi_\rho A^*f_\rho^2A\pi_\rho^*
\;=\;
f_\rho A^*A f_\rho \,+\,
\pi_\rho \big([f_\rho,A]^*f_\rho A+f_\rho A^*[f_\rho ,A]\big)\pi_\rho^*
\;.
$$
Now the first term on the r.h.s. combined with a similar one for $A_\rho A_\rho^*$ and the second term of \eqref{eq-LowBound} leads to a uniform lower bound by $g^2$ on $\Hh_\rho\oplus\Hh_\rho$. The commutators are bounded by $\|[f_\rho ,A]\|\leq \|\widehat{f'_\rho}\|_{L^1}\|[D,A]\|$ (see \cite{BR}) and are dealt with as perturbations, just as the last summand in \eqref{eq-LowBound}. On a technical level, these perturbations are then controlled by \eqref{eq:kappa} and it is a matter of patience to combine all of these estimates to obtain \eqref{eq-LBound}, see \cite{LS2,LS3}.

\vspace{.2cm}

As long as $L_{\kappa,\rho}$ is invertible, its signature is well-defined. The signature is the finite-dimensional equivalent of the $\eta$-invariant. The following result states its stability properties.

\begin{theo}
\label{theo-SigConst}
As long as \eqref {eq:kappa} and \eqref{eq:rho} hold, $\Sig(L_{\kappa,\rho})$ is independent of \(\kappa\) und \(\rho\).
\end{theo}

The proof of Theorem~\ref{theo-SigConst} is again given in \cite{LS2,LS3}, but it merely interpolates between different values of $\kappa$ and $\rho$ for which the gap is known to be open by Theorem~\ref{theo-Gap}. As the gap of $L_{\kappa,\rho}$ remains open during these deformations, the signature clearly cannot change. Now the main result on the spectral localizer can be stated. 

\begin{theo}
\label{theo-SigInd}
Suppose that \eqref{eq:kappa} and \eqref{eq:rho} hold. For an even Fredholm module, also suppose that $D_0$ is normal.  Then for the index pairings $T$ given by \eqref{eq-OddPair} and \eqref{eq-EvenPair}, one has
$$
\Ind (T) \;=\; \frac{1}{2} \;\Sig  \left( L_{\kappa, \rho} \right)
\;.
$$
\end{theo}

As already stated above, a proof of Theorem~\ref{theo-SigInd} is given in \cite{LS2,LS3}, and a new proof in the case of even index pairings is outlined in Section~\ref{sec-SFargument}, and the details are carried out in Section~\ref{sec-Proof}.

\section{Chern numbers and half-signatures}
\label{sec-Chern}

This section shows how to apply the spectral localizer in a concrete situation which appears in the analysis of solid state systems. While the presentation of the mathematical framework is essentially self-contained, it is kept very brief because the reader can consult the monograph \cite{PS} for details and further background informations. Let us start out with a Hamiltonian $H=H^*$ on a Hilbert space $\Hh=\ell^2(\ZM^2)\otimes\CM^L$ over a two-dimensional lattice which is supposed to have a spectral gap at $0$ and to be of short-range, namely if $|n\rangle=|n_1,n_2\rangle$ denotes the ($\CM^L$-vector-valued) state localized at $n=(n_1,n_2)\in\ZM^2$, then $\langle n|H|m\rangle=0$ if $|n-m|>R$ for some $R$ (called the range). Then the so-called Fermi projection $P=\chi(H<0)$ has an index pairing w.r.t. the even Fredholm module specified by $D_0=X_1+\imath X_2$ where $X_1$ and $X_2$ denote the unbounded, self-adjoint position operators on $\Hh$ given by $X_j|n_1,n_2\rangle=n_j\,|n_1,n_2\rangle$. At the origin one can modify $D_0$ in order to make it invertible. The boundedness of $[H,D_0]$ is supposed to hold. The index pairing is defined as in \eqref{eq-EvenPair} with 
$$
F\;=\;\frac{X_1+\imath X_2}{|X_1+\imath X_2|}
\;,
$$
together with the choice $F|0\rangle=|0\rangle$ at the origin. By a well-known index theorem, this index is connected to a Chern number whenever the latter is defined. For this, it is necessary that $H$ is either periodic or at least given by a covariant family $(H_\omega)_{\omega\in\Omega}$ of short-ranged, gapped Hamiltonians indexed by a parameter taken from a compact probability space $\Omega$ equipped with a $\ZM^2$-action $\tau$ and an invariant and ergodic probability measure $\PM$. Covariance means that $U(a)H_\omega U(a)^*=H_{\tau_a\omega}$ for $a\in\ZM^2$ and $U(a)$ the magnetic translations (see \cite{PS} for details). Then also the Fermi projections $P_\omega=\chi(H_\omega<0)$ form a covariant family. Each projection $P_\omega$ leads to a Fredholm operator $T_\omega$ by  \eqref{eq-EvenPair} and thus an index $\Ind(T_\omega)$, but it is known that these indices are $\PM$-almost surely constant. On the other hand, the Chern number is defined as
$$
\Ch(P)
\;=\;-\,2\pi\imath\;
\int \PM(d\omega)\;\Tr\,\langle 0|P_\omega[[X_1,P_\omega],[X_2,P_\omega]]|0\rangle
\;,
$$
whenever
\begin{equation}
\label{eq-LocCond}
\sum_{j=1,2}\int \PM(d\omega)\;\Tr\,\langle 0|\,|[X_j,P_\omega]|^2\,|0\rangle
\;<\;\infty
\;.
\end{equation}
When this condition \eqref{eq-LocCond} holds, it is well-known that the Chern number is essentially equal to the Hall conductance. An index theorem (Corollary~6.3.2 in \cite{PS} or \cite{BES}) shows that the almost sure index $\Ind(T_\omega)$ is equal to $\Ch(P)$. The condition \eqref{eq-LocCond} is called the dynamical localization and is considered here as the mathematical definition of the mobility gap regime \cite{BES}. If $0$ lies in a gap, it definitely holds. Combined with Theorem~\ref{theo-SigInd} one therefore obtains

\begin{coro}
\label{coro-Chern}
Let $H=(H_\omega)_{\omega\in\Omega}$ be a covariant family of short range Hamiltonians on the Hilbert space $\Hh=\ell^2(\ZM^2)\otimes\CM^L$ for which $0$ does not lie in the spectrum. Let the spectral localizer be defined by \eqref{eq-SLEvDef} with $D_0=X_1+\imath X_2$. Then the Chern number of the Fermi projection $P=\chi(H<0)$ is given by
$$
\Ch(P)\;=\; \frac{1}{2} \;\Sig  \left( L_{\kappa, \rho} \right)
\;,
$$
provided that $\kappa$ and $\rho$ are chosen such that \eqref{eq:kappa} and \eqref{eq:rho} hold. 
\end{coro}

Let us note a few remarkable facts about this result with a particular focus on numerical implementation. First of all, the construction of the matrix $L_{\kappa, \rho}$ does not involve any spectral calculus of $H$ (in contrast: calculating the Fermi projection requires a diagonalization of $H$). One merely needs the Hamiltonian in the natural basis of $\Hh=\ell^2(\ZM^2)\otimes\CM^L$. Moreover, the finite volume restriction can then be done either on a discrete square box $[-\rho,\rho]^2\cap\ZM^2$ or a sphere $\{(n_1,n_2)\in\ZM^2\,:\,n_1^2+n_2^2\leq \rho^2\}$. The latter appears in Theorem~\ref{theo-SigInd} and Corollary~\ref{coro-Chern}, but it is straightforward to check that the different geometry does not alter the result \cite{LS2}. A second important point is that it is not necessary to carry out a full spectral analysis of the spectral localizer neither. Merely the signature is needed which can be calculated very efficiently by a block Cholesky decomposition. Finally, let us note that in typical situations the value of $\rho$ does not have to be very large so that only relatively small matrices have to be dealt with. All of this is illustrated in Section~\ref{sec-p+ip} below, where also the spectral localizer in the mobility gap regime \eqref{eq-LocCond} is analyzed numerically.

\vspace{.2cm}

Theorem~\ref{theo-SigInd} can readily be applied to other strong invariants appearing in the analysis of topological insulators so that Corollary~\ref{coro-Chern} should be considered as the two-dimensional case. In dimension one and three (and more generally any odd dimension), one considers chiral Hamiltonians which then have a Fermi unitary of which one can calculate integer-valued (higher) winding numbers. These can again be computed by the spectral localizer, see the discussion in \cite{LS}. Also higher Chern numbers are of relevance. For example, a periodically time-driven three-dimensional system can have a non-vanishing second Chern number which is then the non-linear response coefficient for the magneto-electric effect \cite{PS}. Again this integer number can be calculated as the half-signature of a spectral localizer.

\begin{figure}
\centering
\includegraphics[width=15cm]{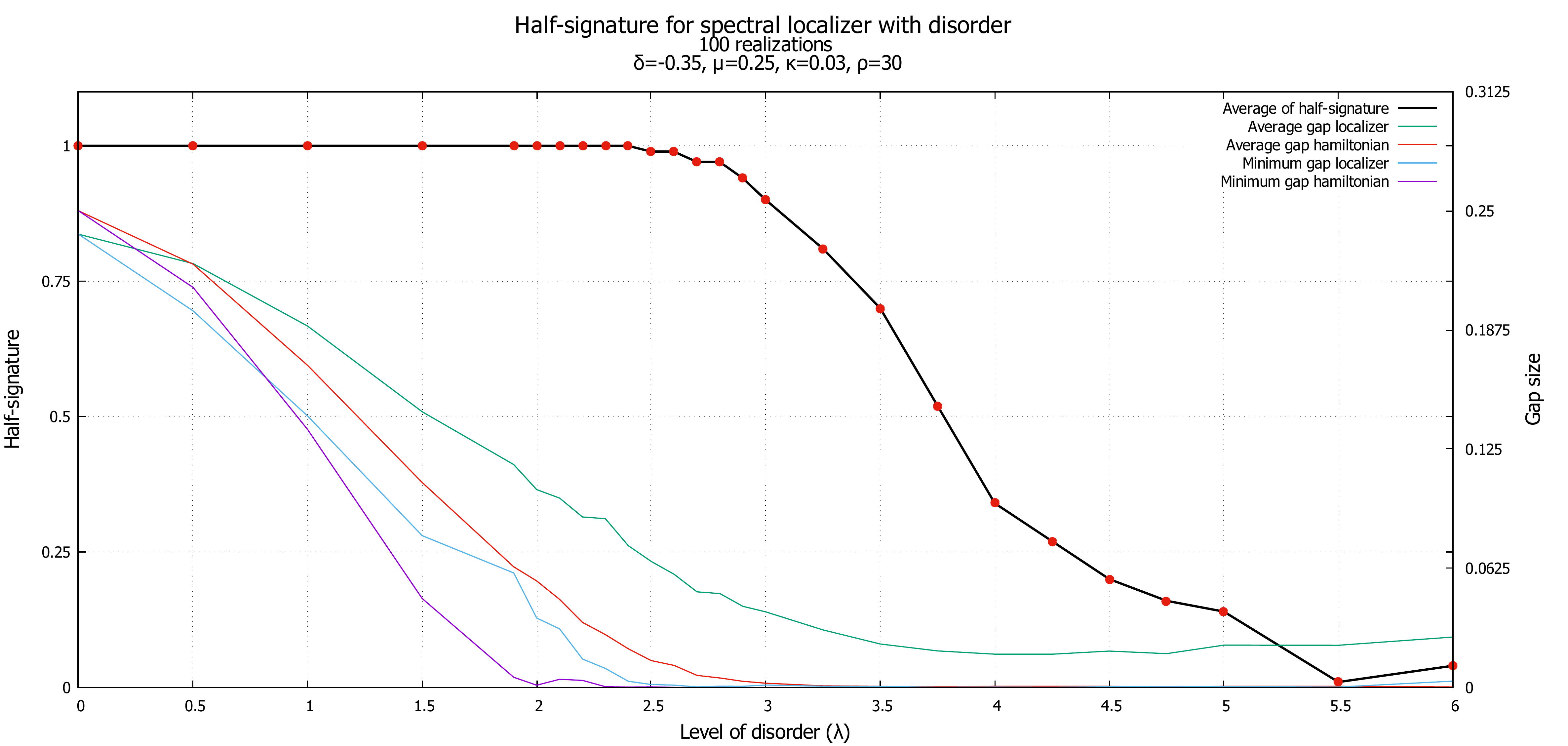}
\caption{Half-signature as well as average and minimum gap sizes of $L_{\kappa,\rho}$ and $H(\lambda)$ (over $100$ samples) as function of $\lambda$. The system size is $\rho=30$ and other parameters as stated.}
\label{fig-HSlambda}
\end{figure}

\section{Numerical results for a dirty $p+ i p$ superconductor}
\label{sec-p+ip}

\begin{figure}
\centering
\includegraphics[width=15cm]{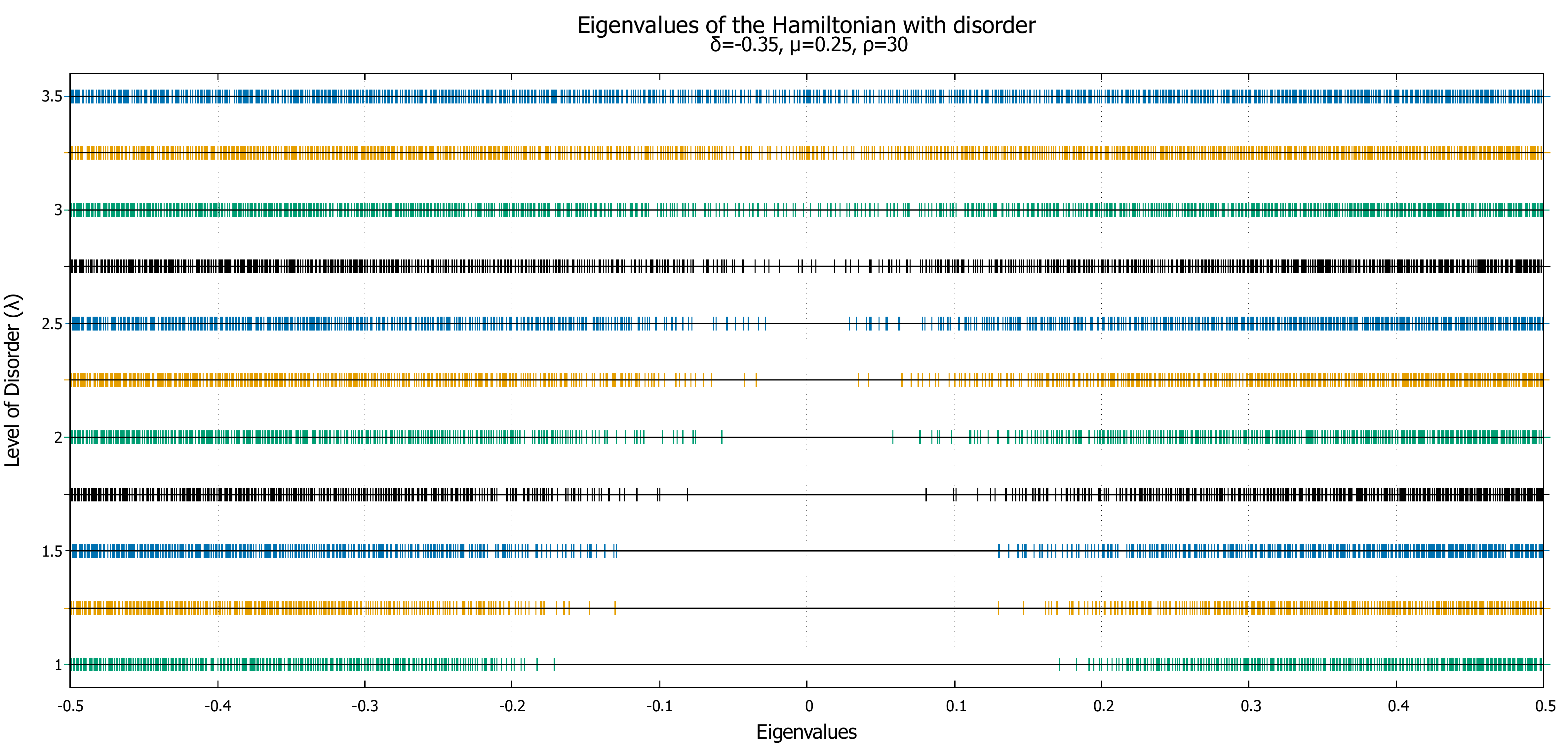}
\caption{Spectrum of $H(\lambda)$ with periodic boundary condition for one realization of the disorder and various values of  $\lambda$.}
\label{fig-SpecH}
\end{figure}

A mean field description of a superconductor leads to Bogoliubov-de Gennes (BdG) Hamiltonian on a particle-hole Hilbert space. For the study of the low-energy behavior it is also sufficient to study a tight-binding BdG Hamiltonian. A well-known topological model of this type is obtained by the $p+ i p$ wave interaction (see \cite{DDS} for references to the physics literature on this model). A periodic (clean) system of this type is described by the BdG Hamiltonian on $\ell^2(\ZM^2,\CM^2)$ of the form
$$
H(0)
\;=\;
\begin{pmatrix}
S_1+S_1^*+S_2+S_2^* -\mu & \delta\big(S_1-S_1^*+\imath(S_2-S_2^*)\big) \\ 
\delta\big(S_1-S_1^*+\imath(S_2-S_2^*)\big)^* & -(S_1+S_1^*+S_2+S_2^* - \mu) 
\end{pmatrix}
\;.
$$
Here $S_1$ and $S_2$ are the shifts on the lattice given by 
$$
S_1|n_1,n_2\rangle\;=\;|n_1+1,n_2\rangle
\;,
\qquad
S_2|n_1,n_2\rangle\;=\;|n_1,n_2+1\rangle
\;.
$$ 
The parameter $\mu\in\RM$ is the chemical potential and $\delta\in\RM$ is the strength of the $p+ip$ pairing potential.  The system becomes a "dirty" superconductor by adding a random potential of the type
$$
V_\omega
\;=\;
\sum_{n\in\ZM^2}v_n\begin{pmatrix} 1 & 0 \\ 0 & -1 \end{pmatrix} \;|n\rangle\langle n|
\;.
$$
Here each realization $\omega=(v_n)_{n\in\ZM^2}$ is a point in the compact Tychonov space $\Omega=[-\frac{1}{2},\frac{1}{2}]^{\ZM^2}$.  Each $v_n$ is drawn independently and identically with a uniform distribution from the interval $[-\frac{1}{2},\frac{1}{2}]$. The product measure $\PM$ on $\Omega$ is then invariant and ergodic w.r.t. to the natural shift action of $\ZM^2$ on $\Omega$. The random BdG Hamiltonian with coupling constant $\lambda\geq 0$ is now: 
\begin{equation}
\label{eq-HamPIP}
H_\omega(\lambda)
\;\;=\;\;
H(0)\,+\,\lambda\,V_\omega
\;.
\end{equation}
Note that it still has the particle-hole symmetry $\sigma_1 \overline{H(\lambda)}\sigma_1=-H(\lambda)$ w.r.t. the first Pauli matrix $\sigma_1=\binom{0\;1}{1\;0}$, but this is not crucial for the following. However, it does lead to a symmetry in the spectrum of $H(\lambda)$ that can also be observed in Figure~\ref{fig-SpecH} (the symmetry is broken for the spectral localizer). This completes the description of the model. It fits into the set-up of Section~\ref{sec-Chern}. In particular, the Chern number $\Ch(P)$ is a well-defined integer number under the mobility gap assumption (which is rigorously known to hold only at the band edges at weak disorder \cite{DDS}). Whenever the central gap is open, Corollary~\ref{coro-Chern} allows to compute $\Ch(P)$ as the half-signature of the localizer. 

\vspace{.2cm}

Before describing the numerical results, let us briefly comment on the methods. The random Hamiltonian \eqref{eq-HamPIP} and associated spectral localizer were generated by a standard random number generator. Of importance is that one uses Dirichlet boundary conditions for the spectral localizer (and the Hamiltonian therein), but the spectra of the finite volume approximations of the Hamiltonian are calculated with periodic boundary conditions. If one uses Dirichlet boundary conditions for the latter, this produces edge states which always close the (bulk) gap and are not the focus of the present study. We simply used an octave code to diagonalize the Hamiltonian and the spectral localizer. The code was run on a Supermicro with two processors Xeon E5-2630 V2 at 2.60ghz and with 128gb RAM. Figure~\ref{fig-HSlambda} needed a few days of CPU time, the others only less than an hour.

\begin{figure}
\centering
\includegraphics[width=15cm]{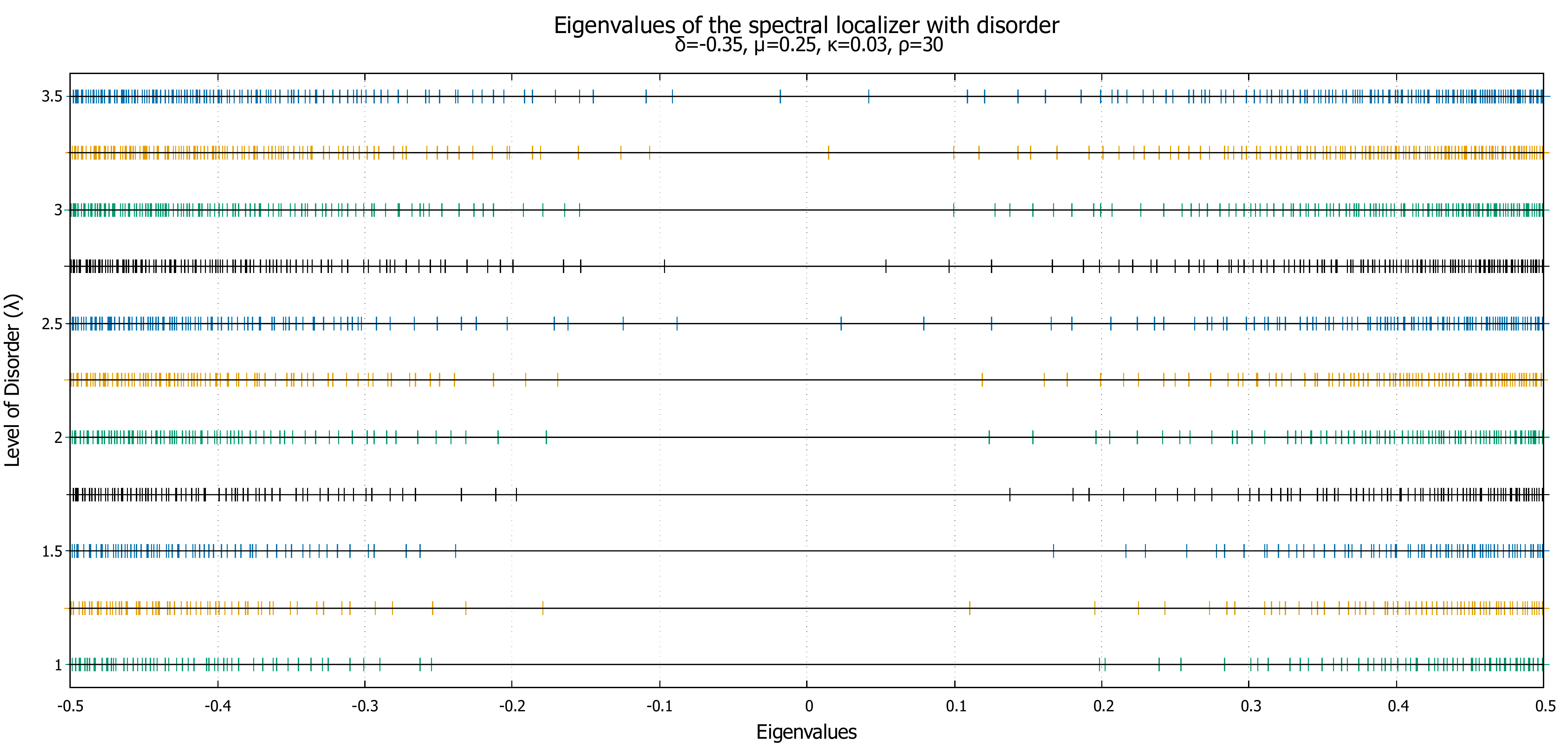}
\caption{Example of the spectrum of the spectral localizer for one realization.}
\label{fig-SLspec}
\end{figure}

\vspace{.2cm}

Let us start by describing the clean Hamiltonian at $\lambda=0$. Its spectrum and Chern numbers can be calculated analytically \cite{DDS}. The Hamiltonian $H(0)$ has a central gap around $0$ except for $(\mu,\delta)$ lying on the coordinate axis. In the four quadrants, the Chern number are $1$ (for $\mu>0$) and $-1$ (for $\mu<0$). We choose a point $(\delta,\mu)=(-0.35,0.25)$ which is well inside the topologically non-trivial phase, but for which the gap of the Hamiltonian $H(0)$ is not too large, namely roughly equal to $0.27$. With these parameters fixed in the following, let us now add the random potential by varying $\lambda$. The results are shown in Figure~\ref{fig-HSlambda}. First of all, one notes that the half-signature is constant for values $\lambda\leq 2.75$. However, one is in the regime of Corollary~\ref{coro-Chern} with an open bulk gap only for $\lambda<2.0$ because Figure~\ref{fig-HSlambda} shows that for $\lambda\geq 2.0$ there are already realizations with a closed gap of $H(\lambda)$. See also Figure~\ref{fig-SpecH} for an illustration of this fact where at least at $\lambda=2.75$ there are eigenvalues very close to $0$. The most remarkable regime in Figure~\ref{fig-HSlambda} is for values of $\lambda\in[2.5,2.75]$. Here the Hamiltonian is not gapped, but expected to be in the mobility gap regime. The half-signature is nevertheless deterministically equal to the non-trivial value $1$. Figure~\ref{fig-SLspec} shows the spectrum of the spectral localizer for one particular realization for various values of $\lambda$ (the same realization as in Figure~\ref{fig-SpecH}). One can clearly see that the central gap of the spectral localizer is open for $\lambda\leq 3$. As $\lambda$ increases further, the averaged half-signature decreases to the topologically non-trivial value $0$, and this also corresponds to a closing gap for the particular realization in Figure~\ref{fig-SLspec}.

\vspace{.2cm}

In conclusion, we believe that these numerical results strongly support the use of the spectral localizer in the mobility gap regime. Clearly further numerical and analytical analysis is needed to gain a better understanding of the spectral localizer in this physically interesting regime.

\section{Spectral flow argument for even index pairings}
\label{sec-SFargument}

As already advertised above, the main mathematical novelty of this paper is a new proof of Theorem~\ref{theo-SigInd} for the case of even pairings. In this section, we will therefore outline the strategy of the argument, deferring detailed proofs of the technical lemmata to Section~\ref{sec-Proof}. Of course, some familiarity with the spectral flow is necessary and the main facts needed here are collected in Appendix~\ref{app-SFReview} for the convenience of the reader. Crucial for the understanding of the following is that $\SF(T_0,T_1)$ is the spectral flow along the straight line path $T_t=(1-t)T_0+tT_1$ connecting two selfadjoint Fredholm operators within the set of Fredholm operators. Further properties of the spectral flow used in the following are the homotopy invariances under homotopies keeping the end points fixed, the invariance under unitary transformations and as well as the concatenation and additivity properties. Finally, the spectral flow is connected to the index of an index pairing $T=PFP+(\one -P)$ by a theorem of Phillips \cite{Phi1}:
$$
\Ind(T)\;=\;\SF \left(F(\one-2P )F^*,\one- 2P \right) 
\;.
$$
As in \cite{LS3}, this is the starting point of the argument. Due to the gap of $H$, one can next deform $\one-2P$ into $H$:
$$
\Ind(T)\;=\;
\SF ( FHF^*,H )
\;.
$$
Furthermore, one can use the additivity of the spectral flow as well as the definition of $\Gamma=\diag(\one,-\one)$ to deduce
\begin{align*}
\Ind (T) 
&\;=\;
\SF \left( \begin{pmatrix} \one  & 0 \\ 0 & F\end{pmatrix}\begin{pmatrix} -H & 0 \\ 0 & H\end{pmatrix}\begin{pmatrix} \one  & 0 \\ 0 & F^*\end{pmatrix},\begin{pmatrix} -H & 0 \\ 0 & H\end{pmatrix}
\right)
\\ &\;=\;\SF \left(\begin{pmatrix} \one  & 0 \\ 0 & F\end{pmatrix} \left(-H \otimes \Gamma\right) \begin{pmatrix} \one  & 0 \\ 0 & F^*\end{pmatrix}, -H \otimes \Gamma\right)
\;.
\end{align*}
Now the following lemma will allow to replace the second argument $-H\otimes \Gamma$ by $L_\kappa$.

\begin{lemma}
\label{lem-SFSLH} 
For $\kappa$ sufficiently small,
$$
\SF(-H\otimes \Gamma,L_\kappa)\;=\;0
\;.
$$
\end{lemma}

\noindent {\bf Proof.} As already stated, most proofs are deferred to Section~\ref{sec-Proof}, but this one is so short and essential that we give it right away. Indeed, one merely checks that the gap does not close along the straight-line path $T_t=-H\otimes\Gamma +t \kappa D$ connecting $-H\otimes \Gamma$ and $L_\kappa$. This follows from
$$
(T_t)^2
\;=\;
\begin{pmatrix} H^2+(t \kappa)^2 |D_0|^2 & t \kappa [H,D_0]^* \\ t \kappa [H,D_0] & H^2+(t \kappa)^2 \|D_0|^2\end{pmatrix}
\;\geq \;
\left(g^2 - \kappa \Vert[D,H \otimes \one_2]\Vert\right) \one 
\;,
$$
for $\kappa$ sufficiently small.
\hfill $\Box$

\vspace{.2cm}

Now one can apply the concatenation property \eqref{eq-SFrule} to deduce 
\begin{equation}
\label{eq-SFLoc}
\Ind(T)
\;=\;
\SF \left( \begin{pmatrix} \one  & 0 \\ 0 & F\end{pmatrix} \left(-H \otimes \Gamma\right) \begin{pmatrix} \one  & 0 \\ 0 & F^*\end{pmatrix},L_{\kappa}\right)
\;.
\end{equation}
Some care is needed at this point because one has to verify that the path given by the concatenation of two straight-line paths can be deformed (within the self-adjoint Fredholm operators) into a straight-line path, but the conditions stated after \eqref{eq-SFrule} are satisfied because the first straight-line path is merely a compact difference and the second is even in the invertibles. A second point is that Lemma~\ref{lem-SFSLH} and thus \eqref{eq-SFLoc} only holds for small $\kappa$. This is, however, sufficient because by Theorem~\ref{theo-SigConst} it is sufficient to prove the claim of Theorem~\ref{theo-SigInd} for $\kappa$ as small as desired and $\rho$ correspondingly large (or larger) so that \eqref{eq:rho} holds.

\vspace{.2cm}

The next point is that one can decouple both $L_\kappa$ and $H\otimes\Gamma$ into their finite volume restrictions and their restrictions to the orthogonal complement $(\Hh_\rho\oplus\Hh_\rho)^\perp=\Ran ( \chi(|D| > \rho))$. To state this fact, let us denote the surjective partial isometry onto the latter space by $\pi_{\rho^c}$ and set $B_{\rho^c}=\pi_{\rho^c} B \pi_{\rho^c}^*$ for any operator $B$.

\begin{lemma}
\label{lem-SLDecop} For $\kappa$ sufficiently small and $\rho$ sufficiently large,
$$
\SF(L_\kappa,L_{\kappa ,\rho} \oplus L_{\kappa ,\rho^c })\;=\;0
\;.
$$
\end{lemma}

\begin{lemma}
\label{lem-HDecop} Let $\kappa$ be sufficiently small and $\rho$ sufficiently large. Supposing that $H_{\rho}$ and $H_{\rho^c }$ are invertible,
$$
\SF(H\otimes \Gamma,(H_{\rho} \oplus H_{\rho^c })\otimes\Gamma)\;=\;0
\;.
$$
\end{lemma}

Both technical proofs are given in Section~\ref{sec-Proof}, but let us give an intuitive argument why these facts are true. First of all, the matrix elements of $L_\kappa$ coupling $L_{\kappa ,\rho}$ to  $L_{\kappa ,\rho^c}$ stem from the operator $H$ and are thus uniformly bounded and local in the sense that they fall off from the boundary. For $\rho$ large, such local and bounded terms are dominated by the operator $D$ which is of the order $\rho$ in this region. Therefore homotopically sending this coupling to $0$ does not modify the low lying spectrum and thus does not lead to a spectral flow, as claimed in Lemma~\ref{lem-SLDecop}. The reason why Lemma~\ref{lem-HDecop} holds is much simpler: the presence of $\Gamma$ assures that tuning down the coupling elements of $H$ connecting $\Hh_\rho$ and $\Hh_{\rho^c}$ leads to as much spectral flow upwards as downwards. The invertibility condition is merely imposed to avoid ambiguities in the definition of the spectral flow. It will also be shown in Section~\ref{sec-Proof} that this can readily be achieved by an arbitrarily small and compact perturbation of $H$. Hence one can assume this to hold in the following. Applying these two lemmas and using again the unitary invariance and additivity \eqref{eq-SFrule} of the spectral flow now implies (note that all straight-line paths involved  only consist of adding compact operators so that the conditions for \eqref{eq-SFrule} are indeed satisfied):
\begin{align}
\Ind(T)
\;=\;
& 
\;\SF \left( \begin{pmatrix} \one _{\rho} & 0 \\ 0 & F_{\rho}\end{pmatrix} \left(-H_{\rho}\otimes \Gamma\right) \begin{pmatrix} \one _{\rho} & 0 \\ 0 & F_{\rho}^*\end{pmatrix},L_{\kappa ,\rho }\right)
\nonumber\\
& 
\;\,+\;
\SF \left(\begin{pmatrix} \one _{\rho^c} & 0 \\ 0 & F_{\rho^c}\end{pmatrix} \left(-H_{\rho^c} \otimes \Gamma\right) \begin{pmatrix} \one _{\rho^c} & 0 \\ 0 & F_{\rho^c}^*\end{pmatrix}, L_{\kappa ,\rho^c }\right)
\;.
\label{eq-IndSFLast}
\end{align}
%

\begin{lemma}
\label{lem-OuterSF} 
The spectral flow of the second summand in \eqref{eq-IndSFLast} vanishes.
\end{lemma}

Hence $\Ind(T)$ is merely given by the first summand in \eqref{eq-IndSFLast}. This is the spectral flow between two finite dimensional selfadjoint matrices, and as such given as half the difference of the signatures of these matrices, notably 
\begin{align*}
\Ind(T) 
&\;=\;\frac{1}{2} \left(\Sig \left(L_{\kappa ,\rho }\right)\;-\;
\Sig \left(\begin{pmatrix} \one _{\rho} & 0 \\ 0 & F_{\rho}\end{pmatrix} \left( -H_{\rho}\otimes \Gamma\right) \begin{pmatrix} \one _{\rho} & 0 \\ 0 & F_{\rho}^*\end{pmatrix}\right) \right) 
\\ &
\;=\;
\frac{1}{2} \big(\Sig \left(L_{\kappa ,\rho }\right) \,+\,\Sig \left( H_{\rho}\otimes \Gamma\right) \big)
\\ &
\;=\;
\frac{1}{2} \;\Sig \left(L_{\kappa ,\rho }\right)
\;.
\end{align*}
This concludes the proof of Theorem~\ref{theo-SigInd} for the case of even pairings.

\section{Details of the spectral flow proof}
\label{sec-Proof}

This section contains the proofs of the lemmata of Section~\ref{sec-SFargument} and further facts needed there.

\vspace{.2cm}

\noindent {\bf Proof} of Lemma~\ref{lem-SLDecop}. Let us set
$$
L_\kappa(t) 
\;=\; 
L_{\kappa ,\rho} \oplus L_{\kappa ,\rho^c } 
\;-\; t \begin{pmatrix} 0 & \pi_\rho \left(H \otimes \Gamma\right) \pi_{\rho^c}^* \\ \pi_{\rho^c} \left(H \otimes \Gamma\right) \pi_\rho^* & 0\end{pmatrix}
\;.
$$
Then $L_\kappa(1)=L_\kappa$ and $L_\kappa(0)=L_{\kappa ,\rho} \oplus L_{\kappa ,\rho^c }$. Because $\pi_\rho$ is of finite range, the second summand on the r.h.s. is compact. It will be shown that this path is in the invertible (Fredholm) operators and this then implies the claim. First let us note that the operator $L_{\kappa ,\rho^c}$ is invertible for $\rho$ sufficiently large because, using \eqref{eq:kappa}, one has
\begin{align*}
(L_{\kappa ,\rho^c })^2 
&
\;=\; 
\begin{pmatrix} H_{\rho^c}^2+\kappa^2 \left|D_{0,\rho^c}\right|^2 & \kappa \left[D_{0,\rho^c},H_{\rho^c}\right]^* \\ \kappa \left[D_{0,\rho^c},H_{\rho^c}\right] & H_{\rho^c}^2+\kappa^2 \left|D_{0,\rho^c}\right|^2\end{pmatrix}
\\&
\;\geq \;\left(\kappa^2 \rho^2 - \kappa \left\Vert\left[D,H \otimes \one\right]\right\Vert\right) \one _{\rho^c}
\\&\;\geq \;\left(\kappa^2 \rho^2 - \frac{g^3}{12 \left\Vert H \right\Vert}\right) \one _{\rho^c}
\\&\;\geq\; \left(\kappa^2 \rho^2 - \frac{g^2}{12}\right) \one _{\rho^c}
\\&\;\geq\; \left(\kappa^2 \rho^2 - \frac{\kappa^2 \rho^2}{48}\right) \one _{\rho^c}
\\&\;\geq\; \frac{1}{2}\, \kappa^2 \rho^2 \,\one _{\rho^c}
\;.
\end{align*}
Introducing the invertible operator $\widetilde L=|L_{\kappa ,\rho } \oplus L_{\kappa ,\rho^c }|^\frac{1}{2}$, one now has
\begin{equation*}
L_\kappa(t)
\;=\;
\widetilde L \left( S - t \begin{pmatrix} 0 & \left|L_{\kappa ,\rho }\right|^{-\frac{1}{2}}\pi_\rho \left(H \otimes \Gamma\right) \pi_{\rho^c}^*\left|L_{\kappa ,\rho^c }\right|^{-\frac{1}{2}} \\ \left|L_{\kappa ,\rho^c }\right|^{-\frac{1}{2}} \pi_{\rho^c} \left(H \otimes \Gamma\right) \pi_\rho^* \left|L_{\kappa ,\rho }\right|^{-\frac{1}{2}} & 0\end{pmatrix} \right) \widetilde L
\end{equation*}
where matrix is w.r.t. the decomposition $\mathcal H \oplus \mathcal{H} = (\mathcal H \oplus \mathcal{H})_\rho \oplus (\mathcal H \oplus \mathcal{H})_{\rho^c}$ and $S$ is a selfadjoint unitary which is also diagonal in this grading. As now
$$
\Big\| 
|L_{\kappa ,\rho }|^{-\frac{1}{2}}\pi_\rho \left(H \otimes \Gamma\right) \pi_{\rho^c}^*|L_{\kappa ,\rho^c }|^{-\frac{1}{2}}\Big\| 
\;\leq\; 
\sqrt {\frac{2}{g}} \;\| H \|\; \sqrt {\frac{\sqrt 2}{\kappa \rho}} 
\;=\; 
\frac{C}{\sqrt {\kappa \rho}}
\;,
$$
the invertibility of $L_\kappa(t)$ follows for sufficiently large $\rho$. 
\hfill $\Box$

\vspace{.2cm}

Concerning the proof of Lemma~\ref{lem-HDecop}, namely that $\SF(H\otimes \Gamma,(H_{\rho} \oplus H_{\rho^c })\otimes\Gamma)=0$, it was already stated above that this results from the spectral doubling $\sigma(H_t\otimes \Gamma)=\sigma(H_t)\cup(-\sigma(H_t))$ along the path associated to $H_t=(1-t) H+t\,H_{\rho} \oplus H_{\rho^c }$. To avoid ambiguities, it is best to assure that the end point $H_1=H_{\rho} \oplus H_{\rho^c }$ is also invertible (see the hypothesis in Lemma~\ref{lem-HDecop}). This can be achieved by the following lemma.

\begin{lemma}
\label{lem-HDecop2} For any $H$ and $\rho$, there exists a selfadjoint $\widetilde{H}$ with 
\vspace{.1cm}

\noindent {\rm (i)} $H-\widetilde{H}$ is of finite rank,

\noindent {\rm (ii)} $H-\widetilde{H}$ is of arbitrarily small norm,

\noindent {\rm (iii)} $\widetilde{H}_{\rho}$ and $\widetilde{H}_{\rho^c }$ are invertible.
\end{lemma}

\noindent {\bf Proof.} Recall that any neighborhood of a self-adjoint Fredholm operator (and more generally of a Fredholm operator with vanishing index) contains an invertible self-adjoint operator (just add a small finite rank perturbation on the finite dimensional kernel). This can be applied to both summands of $H_\rho \oplus H_{\rho^c}$ separately, and then combined to show the  claim.
\hfill $\Box$

\vspace{.2cm}

Lemma~\ref{lem-HDecop2} can be used to replace $H$ by $\widetilde{H}$ at any stage of the argument described in Section~\ref{sec-SFargument} because the (arbitrarily) small paths from $H$ to $\widetilde{H}$ never lead to spectral flow. In particular, this lifting of the kernels of ${H}_{\rho}$ and ${H}_{\rho^c }$ can be done at the very beginning.  We will suppress the distinction of $H$ from $\widetilde{H}$ and can thus tacitly assume that ${H}_{\rho}$ and ${H}_{\rho^c }$ are both invertible from now. Now essentially only remains the

\vspace{.2cm}

\noindent {\bf Proof} of Lemma~\ref{lem-OuterSF}. Let us use the abbreviation 
$$
\SF_c
\;=\;
\SF \left( L_{\kappa ,\rho^c },\begin{pmatrix} \one _{\rho^c} & 0 \\ 0 & F_{\rho^c}\end{pmatrix} \left(-H_{\rho^c} \otimes \Gamma\right) \begin{pmatrix} \one _{\rho^c} & 0 \\ 0 & F_{\rho^c}^*\end{pmatrix}\right)
\;.
$$
Hence the aim is to show $\SF_c=0$. The first step is to show that
\begin{equation}
\label{eq-Intermed}
\SF_c
\;=\;
\SF \left( L_{\kappa ,\rho^c },\begin{pmatrix} \one _{\rho^c} & 0 \\ 0 & F_{\rho^c}\end{pmatrix} \begin{pmatrix}  -H_{\rho^c} & \kappa \rho \\ \kappa \rho &  H_{\rho^c}\end{pmatrix} \begin{pmatrix} \one _{\rho^c} & 0 \\ 0 & F_{\rho^c}^*\end{pmatrix}\right)
\;.
\end{equation}
This seems to follow immediately from the standard bound
\begin{align}
\label{eq-DiracGap}
\begin{pmatrix} -H_{\rho^c} & \lambda\kappa\rho\,\one _{\rho^c} \\ \lambda\kappa\rho\,\one _{\rho^c} & H_{\rho^c}\end{pmatrix}^2 
&
\;=\;\begin{pmatrix} H_{\rho^c}^2+\lambda^2\kappa^2\rho^2\,\one _{\rho^c} & 0 \\ 0 & H_{\rho^c}^2+\lambda^2\kappa^2\rho^2\,\one _{\rho^c} \end{pmatrix}
\nonumber
\\
&
\;\geq \;\big(\| H_{\rho^c}^{-1} \|^{-2} + \lambda^2\kappa^2\rho^2\big)\one _{\rho^c}
\end{align}
for $\lambda\in[0,1]$ and the fact that $H_{\rho^c}$ is invertible because then there is no extra spectral flow on the line segment connecting $H_{\rho^c} \otimes \Gamma$ to the middle matrix in the second argument of \eqref{eq-Intermed}, but there is a caveat here because one still needs to verify that the two parameter family
$$
A(t,\lambda)
\;=\; 
t\, L_{\kappa ,\rho^c } 
\;+\; 
(1-t) \begin{pmatrix} \one _{\rho^c} & 0 \\ 0 & F_{\rho^c}\end{pmatrix} \begin{pmatrix}  -H_{\rho^c} & \lambda \kappa \rho \\ \lambda \kappa \rho & H_{\rho^c}\end{pmatrix} \begin{pmatrix} \one _{\rho^c} & 0 \\ 0 & F_{\rho^c}^*\end{pmatrix}
$$
lies in the selfadjoint Fredholms so that the additivity rule \eqref{eq-SFrule} applies.  This Fredholm property is not altered if one adds finite rank operators such as the complementing piece of $\Hh_\rho\oplus\Hh_\rho$, and even compact operators such as $\pi_\rho H\pi_{\rho^c}^*$ and $[H,F]$. Hence the Fredholm property of $A(t,\lambda)$ is equivalent to the Fredholm property of 
\begin{align*}
B\left(t,\lambda\right) 
&\;=\; t\, L_{\kappa} \;+\; \left(1-t\right) \begin{pmatrix} -H & \lambda \kappa \rho F^* \\ \lambda \kappa \rho F & H\end{pmatrix}
\\
&
\;=\; \begin{pmatrix} -H & t \kappa D_0^* + \left(1-t\right) \lambda \kappa \rho F^*  \\ t \kappa D_0 + \left(1-t\right) \lambda \kappa \rho F & H\end{pmatrix}
\;.
\end{align*}
Now the off-diagonal entries combined are a function of $D$. This function involves an indicator function, but it can be replaced by a smooth function, up to a compact perturbation. This shall be explored next. Let $s_1: \mathbb{R} \to [-1,1]$ be a smooth function with $s_1|_{\left(-\infty,-1\right]} = -1$, $s_1|_{\left[1,\infty\right)} = 1$  and $\|\widehat {s'_1}\|_{L^1\left(\mathbb{R}\right)} =2$. Such functions are explicitly constructed in Lemma~4 of \cite{LS} where it is also shown that $s_\rho\left(x\right)=s_1(\frac{x}{\rho})$ then satisfies
$$
\left\|\left[s_\rho\left(D\right),H \otimes \one\right]\right\| 
\;\leq \;
2\, \rho^{-1} \,\left\|\left[D,H \otimes \one\right]\right\|
\;.
$$
Now introduce the smoothening $\widetilde{F}$ of $F$ by
$$
s_\rho\left(D\right)\;=\;\begin{pmatrix} 0 & \widetilde F^*  \\ \widetilde F & 0\end{pmatrix}
\;.
$$
Then $\|[\widetilde F,H]\| \leq 2 \rho^{-1} \|[D,H \otimes \one]\|$ and still ${\widetilde F=\widetilde F_\rho \oplus \widetilde F_{\rho^c}}$. Furthermore, $\widetilde F_{\rho^c} = F_{\rho^c}$ so that $\widetilde F - F$ is compact. Hence $B\left(t,\lambda\right) $ is Fredholm if and only if 
$$
C\left(t,\lambda\right)\;=\; \begin{pmatrix} -H & t \kappa D_0^* + \left(1-t\right) \lambda \kappa \rho \widetilde F^*  \\ t \kappa D_0 + \left(1-t\right) \lambda \kappa \rho \widetilde F & H\end{pmatrix}
$$
is Fredholm. Now
\begin{align*}
C\left(t,\lambda\right)^2 
&\;=\; \begin{pmatrix} H^2+|t \kappa D_0 + (1-t) \lambda \kappa \rho \widetilde F|^2 & [H,t \kappa D_0 + (1-t) \lambda \kappa \rho \widetilde F]^*  \\  
[H,t \kappa D_0 + (1-t) \lambda \kappa \rho \widetilde F] & H^2+|t \kappa D_0 + (1-t) \lambda \kappa \rho \widetilde F|^2\end{pmatrix}
\\
&
\;\geq\; 
g^2
\,-\,t \kappa \|[D,H \otimes \one]\| \,-\, (1-t) \lambda \kappa \rho\, \|[\widetilde F,H]\| 
\\
&
\;\geq \;
g^2 \,-\, \kappa \|[D,H \otimes \one]\| \,-\, \kappa \rho  2 \rho^{-1} \|[D,H \otimes \one]\|
\\
&
\;=\; g^2 \,-\, 3 \kappa \|[D,H \otimes \one]\|
\;.
\end{align*}
This last expression is strictly positive for $\kappa$ sufficiently small. Hence then $C(t,\lambda)$ is invertible and thus a Fredholm operator.  By now, a formal proof of \eqref{eq-Intermed} is completed. Let us next multiply out:
$$
\SF_c
\;=\;
\SF \left( L_{\kappa ,\rho^c },\begin{pmatrix} -H_{\rho^c} & \kappa \rho F_{\rho^c}^* \\ \kappa \rho F_{\rho^c} & F_{\rho^c} H_{\rho^c} F_{\rho^c}^*\end{pmatrix}\right)
\;.
$$
Next $F_{\rho^c} H_{\rho^c} F_{\rho^c}^*- H_{\rho^c}$ is compact with operator norm bounded by $2\|H\|$.  Thus using \eqref{eq-DiracGap} for $\lambda=1$,
$$
\left\| \begin{pmatrix} -H_{\rho^c} & \kappa \rho F_{\rho^c}^* \\ \kappa \rho F_{\rho^c} & H_{\rho^c}\end{pmatrix}^{-1} \right\|^{-1} 
\;\geq\; 
 \kappa \rho \,-\, 2\, \| H \| 
\;,
$$
which is thus positive for $\rho$ sufficiently large. As the path connecting $F_{\rho^c} H_{\rho^c} F_{\rho^c}^*$ to $ H_{\rho^c}$ is in the compacts,
$$
\SF_c
\;=\;
\SF \left( L_{\kappa ,\rho^c },\begin{pmatrix} -H_{\rho^c} & \kappa \rho F_{\rho^c}^* \\ \kappa \rho F_{\rho^c} & H_{\rho^c}\end{pmatrix}\right)
\;.
$$
It only remains to show that this last expression vanishes. For this, it is shown that the straight-line path
$$
D(\lambda)
\;=\;
\begin{pmatrix} -H_{\rho^c} & \kappa \left(\left(1-\lambda\right) \rho F_{\rho^c}^* + \lambda {D_0^*}_{\rho^c}\right) \\ \kappa \left(\left(1-\lambda\right) \rho F_{\rho^c} + \lambda D_{0,\rho^c}\right) & H_{\rho^c}\end{pmatrix}
$$
connecting the two arguments lies in the invertible operators. Indeed, using 
$$
\left|
(1-\lambda)\rho F_{\rho^c}\,+\,\lambda\,D_{0,\rho^c}
\right|
\;\geq\;
\rho^2\,\one_{\rho^c}
\;,
$$
one finds
\begin{align*}
D(\lambda)^2
&
\;=\;
\begin{pmatrix} (H_{\rho^c})^2+\kappa^2 \left|\left(1-\lambda\right) \rho F_{\rho^c} + \lambda D_{0,\rho^c}\right|^2 & \kappa \left[H_{\rho^c},\left(1-\lambda\right) \rho F_{\rho^c} + \lambda D_{0,\rho^c}\right]^* \\ \kappa \left[H_{\rho^c},\left(1-\lambda\right) \rho F_{\rho^c} + \lambda D_{0,\rho^c}\right] & (H_{\rho^c})^2+\kappa^2 \left|\left(1-\lambda\right) \rho F^*_{\rho^c} + \lambda {D_0}^*_{\rho^c}\right|^2\end{pmatrix}
\\ 
&
\;\geq \;
\big(\left(\kappa \rho\right)^2-\kappa \rho \left\Vert\left[F,H\right]\right\Vert - \kappa \left\Vert\left[D,H \otimes \one\right]\right\Vert\big) \one _{\rho^c}
\\ &\;=\; \kappa \big(\rho \left(\kappa \rho - \left\Vert[F,H]\right\Vert\right) - \left\Vert\left[D,H \otimes \one\right]\right\Vert \big)\one _{\rho^c}
\;,
\end{align*}
which is strictly positive for $\rho$ sufficiently large.
\hfill $\Box$

\vspace{.3cm}

\noindent {\bf Acknowledgements:} This work was partially supported by the DFG.

\appendix

\section{Properties of the spectral flow}
\label{app-SFReview}

This appendix collects the relevant information from \cite{Phi1,Phi2,DS2} on the spectral flow. Let $t\in[0,1]\mapsto T_t$ be a continuous path of bounded selfadjoint Fredholm operators. Then there is an $\epsilon>0$ such that uniformly in $t\in[0,1]$ the operator $T_t$ has only discrete spectrum  (isolated  eigenvalues of finite multiplicity) in $(-\epsilon,\epsilon)$. Intuitively, the spectral flow is then the number of eigenvalues moving past $0$ in the positive direction minus the number of those eigenvalues moving past $0$ in the negative direction. In \cite{Phi1}, Phillips gives a careful definition of the spectral flow that is not spelled out here. However, let us spell out the main properties of the spectral flow:
\begin{itemize}

\item[{\rm (i)}] (Homotopy invariance) Let $s\in[0,1]\mapsto T_t(s)$ be a homotopy of paths with fixed end points $T_0(s)$ and $T_1(s)$. Then
$$
\SF(t\in[0,1]\mapsto T_t(0))
\;=\;
\SF(t\in[0,1]\mapsto T_t(1))
\;.
$$

\item[{\rm (ii)}] (Concatenation)  For paths $t\in[0,1]\mapsto T_t$ and $t\in[1,2]\mapsto T_t$,
$$
\SF(t\in[0,1]\mapsto T_t)
\;+\;
\SF(t\in[1,2]\mapsto T_t)
\;=\;
\SF(t\in[0,2]\mapsto T_t)
\;.
$$
\item[{\rm (iii)}] (Unitary invariance) For any path $t\in [0,1]\mapsto U_t$ of unitaries,
$$
\SF(t\in[0,1]\mapsto U_t^*T_tU_t)
\;=\;
\SF(t\in[0,1]\mapsto T_t)
\;.
$$

\item[{\rm (iv)}] (Additivity)  For paths $t\in[0,1]\mapsto T_t$ and $t\in[0,1]\mapsto T'_t$,
$$
\SF(t\in[0,1]\mapsto T_t\oplus T'_t)
\;=\;
\SF(t\in[0,1]\mapsto T_t)
\;+\;
\SF(t\in[0,1]\mapsto T'_t)
\;.
$$

\item[{\rm (v)}] For a path $t\in[0,1]\mapsto T_t$ with $0$ not in the spectrum $\sigma(T_t)$ of $T_t$ for all $t\in[0,1]$,
$$
\SF(t\in[0,1]\mapsto T_t)
\;=\;
0
\;.
$$
\end{itemize}

For the case that the path is given by the straight line between its endpoints, we will also use the shorter notation
$$
\SF (T_0,T_1) \;=\; \SF \big( t\in [0,1]  \mapsto T_t\;=\;(1-t) T_0 + t T_1 \big)
\;.
$$
Given $\SF (T_0,T_1)$ and $\SF (T_1,T_2)$, one has due to (i) and (iii) that
\begin{equation}
\label{eq-SFrule}
\SF (T_0,T_1) \;+\; \SF (T_1,T_2) \;=\; \SF (T_0,T_2)\;,
\end{equation}
provided that all operators of the form $(1-t) T_0 + t (T_1 + \lambda (T_2-T_1))$ are Fredholm. As $(1-t) T_0 + t T_1$ and $(1-\lambda) T_1 + \lambda T_2$ are Fredholm, this is, in particular, the case when either $T_1-T_0$ or $T_2-T_1$ is compact, because 
\begin{align*}
(1-t) T_0 \,+\, t \big(T_1\, +\, \lambda (T_2-T_1)\big) 
&
\;=\; \big((1-t) T_0 \,+\, t T_1\big) \,+ \,t \lambda (T_2-T_1)
\\ &\;=\;(1-t) (T_0-T_1) \,+\, \big((1-t \lambda)T_1 \,+\, t \lambda T_2\big)
\;.
\end{align*}

The starting point of the spectral flow argument is the following fundamental relation between index pairings and spectral flow. It goes back to Phillips \cite{Phi1}. A proof based on homotopy invariance is given in \cite{DS2}.

\begin{theo}
Let $P$ be a projection and $F$ a unitary operator such that $[F,P]$ is compact. Then $PFP+\one-P$ is a Fredholm operator and its index satisfies
\begin{equation}
\label{eq-SFInd}
\Ind(PFP+\one-P)
\;=\;
\SF(F(\one-2P)F^*,\one-2P)
\;.
\end{equation}
\end{theo}

Finally let us add a comment on the spectral flow of paths $t\in[0,1]\mapsto D_t$ of unbounded selfadjoint Fredholm operators. One then obtains a path $t\in[0,1]\mapsto T_t=\tanh(D_t)$ of bounded selfadjoint Fredholm operators and can use its spectral flow to define the spectral flow of $t\in[0,1]\mapsto D_t$. Instead of $\tanh$ any increasing smooth function $f$ with $f(0)=0$ and $f'(0)>0$ can be used. All of the above properties naturally transpose to the unbounded case.


\end{document}